\documentclass[runningheads]{llncs}

\usepackage{graphicx}
\usepackage{subfiles}
\usepackage{cmap}
\usepackage{svg}
\usepackage{verbatim}
\usepackage{xcolor,colortbl}
\definecolor{LightCyan}{rgb}{0.88,1,1}

\usepackage{amsmath} 
\usepackage{amsfonts}
\usepackage{hyperref} 
\usepackage{comment} 
\usepackage{cite} 

\hypersetup{colorlinks=true,citecolor=blue,urlcolor=blue,linkcolor=blue}

\hypersetup{final}
\begin{document}
\title{Citation network applications in a scientific co-authorship recommender system}

\titlerunning{Citation network applications in a scientific recommender system}

\author{Vladislav Tishin \inst{1, 2}, Artyom Sosedka\inst{1, 2}, Peter Ibragimov\inst{2}, \\ \and Vadim Porvatov\inst{1, 2}*}
\authorrunning{Tishin et al.}

\institute{Sberbank, 117997 Moscow, Russia, \and
National University of Science and Technology ``MISIS'', 119991 Moscow, Russia,
\email{*eighonet@gmail.com}\\
\url{}}

\maketitle              
\begin{abstract}

The problem of co-authors selection in the area of scientific collaborations might be a daunting one. In this paper, we propose a new pipeline that effectively utilizes citation data in the link prediction task on the co-authorship network. In particular, we explore the capabilities of a recommender system based on data aggregation strategies on different graphs.
Since graph neural networks proved their efficiency on a wide range of tasks related to recommendation systems, we leverage them as a relevant method for the forecasting of potential collaborations in the scientific community. 

\keywords{Graph machine learning \and 
        Neural networks \and 
        Recommender systems \and 
        Social graphs}
\end{abstract}
\section{Introduction}

Since the advent of scientific communities, there has been a high demand in the area of collaboration recommendations. Due to the complex nature of interconnections between researchers, this domain has not reached a successful automation for a long time.

According to the underlying graph structure of collaboration networks, we propose to use recently emerged graph neural networks (GNN) to efficiently predict research cooperation between scientists. This branch of machine learning has readily proved its outstanding performance in a wide range of areas related to recommender systems \cite{graph_recommender}. Such algorithms as Node2Vec \cite{node2vec}, Attri2Vec \cite{a2v}, and GraphSAGE \cite{GraphSAGE} can be trained to capture structural features of co-authorship network. Embeddings produced by these methods can be effectively applied to the different forecasting tasks including prediction of network connections as well \cite{liu2010link}.  

Graph neural networks allow us not only to boost performance in straightforward link prediction task on co-authorship network, but to improve the quality on such a forecasting challenge via aggregation of additional information from the citation graph. Future development of the discussed pipeline can lead to the simplification of the collaboration assessment process for the R$\&$D team management.

\section{Related work}

Recommender systems for scientific communities have a long history of development.
Early approaches in this area \cite{sie2012whom, recommender2017} were based on deterministic network information, which lacked the ability to represent complex features of graph data.

Learning algorithms in the area of link prediction resolved a variety of issues related to capture of graph intricacies. The explicit examples of such models are local random walk \cite{liu2010link} and local naive Bayes \cite{liu2011link}. 

Various metrics such as content similarity LDAcosin \cite{chuan2018link} were also applied in this field. Another popular method \cite{makarov_recsys} leverages linear regression on feature vectors of nodes and set of graph measures \cite{makarov2017scientific}. However, implementation of a learning feature extractor instead of deterministic metrics would significantly boost the performance of such pipelines. Applying this idea to the collaboration network structures, usage of graph neural networks becomes native \cite{zhang2018link}. 

Despite the promising results achieved by graph neural networks, different techniques could be implemented to further increase their efficiency. Those methods include the alteration of graph topology \cite{singh2021edge} and the usage of task-independent techniques like Node2Vec in order to create initial node representations which serve as better inputs for GNNs \cite{gupta2021integrating}.

\section{Data}

We use classic HEP-TH dataset \cite{hep-th} as the subject for further development and extension. It consists of citation and co-authorship graphs obtained from the arXiv papers published between January 1993 and April 2003 in the area of high energy physics theory. Unfortunately, there is no connection between these two parts of the dataset (authors’ IDs were not provided in the citation network) which makes its initial second part worthless for our purposes. 

To our best knowledge, the potential of the HEP-TH citation network was never explicitly revealed. Due to the presence of highly useful paper metadata (such as author lists or abstracts), the range of the dataset usage can be significantly extended. In order to complete our current research, we perform processing of the citation graph metadata aiming to restore the corresponding co-authorship network. To preserve homogeneous nature of the reconstructing graph, we discarded from the citation graph all anonymous papers.

Along with the information about authors and abstracts, HEP-TH involves a journal reference field with the publisher output information. The presence of such data provides us an opportunity to extract ISSNs of indexed publications and further parse scientific metrics (quartile, h-index, and impact factor) from the "SCImago Journal \& Country Rank" website. 

\section{Methods}

\sloppy
Explored architecture consists of two subsequently applied graph neural networks. The first model generates vector representations of the publications according to their annotations and the structure of the citation network. As its input we leverage the directed citation graph $G(V, A, X)$, where $V = \{ v_1, v_2, \dots, v_n\}$ corresponds to the set of graph nodes (articles), $A: n \times n \rightarrow \{0, 1\}$  is the adjacency matrix (each edge encodes citation between two papers), and $X: n \times d \rightarrow \textbf{R}$ denotes the  matrix of node features (vectorized abstracts via pre-trained FastText \cite{fasttext}). 

We perform a set of computational experiments using previously addressed unsupervised methods GraphSAGE, Node2Vec, and Attri2Vec. In the following, we briefly discuss each of them to clarify their usage as the feature extractors.

\textbf{GraphSAGE}. This method aggregates information about the set of neighbor nodes $N(v)$ in order to produce embedding of node $v$ 

\begin{equation}
h^{l+1}_v = \sigma(W^l \cdot \operatorname{CONCAT}(h^l_v, h^{l+1}_{N(v)})),
\end{equation}
where $l + 1$ is the current number of a convolutional layer, $h^0_v$ = $x_v$, $W^l$ is the matrix of learning parameters, and $h^{l+1}_{N(v)}$ can be extracted by different aggregation functions like Max Pooling or Mean aggregator. 

\textbf{Node2Vec}. This algorithm generates sequences of nodes via second-order random walks and utilizes them as the input data for a skip-gram model. The skip-gram generates pairs from input and context nodes in order to cast them to the feedforward neural network. Its weights can be used as the desired node embeddings as a result of the following function optimization:
\begin{equation}
\max _{f} \sum_{u \in V}\left[-\log Z_{u}+\sum_{n_{i} \in N(u)} f\left(n_{i}\right) \cdot f(u)\right],
\end{equation}
where $Z_{u}=\sum_{v \in V} \exp (f(u) \cdot f(v))$ is per-node partition function and f($\cdot$) corresponds to the mapping function.

\textbf{Attri2Vec}. 
The last considered model uses the image $f(x_i)$ of node $v_i$ with feature vector $x_i$ in the new attribute subspace to predict its context nodes. 
In this method, the task is to solve the joint optimization problem

\begin{equation}
\min _{W^{in}, W^{\text {out }}} -\sum_{i=1}^{|V|} \sum_{j=1}^{|V|} n\left(v_{i}, v_{j}\right) \log \frac{\exp \left(f\left(v_{i}\right) \cdot w_{j}^{\text {out }}\right)}{\sum_{k=1}^{|V|} \exp \left(f\left(v_{i}\right) \cdot w_{k}^{\text {out }}\right)},
\end{equation}
where $n(v_i, v_j)$ is the number of times that $v_j$ occurs in $v_i$ context within t-window size in the generated set of random walks, $W_{\text{in}}$ is the weight matrix from the input layer to hidden layer and $W_{\text{out}}$ is the weight matrix from the hidden
layer to the output layer.

Link prediction step follows after the publications embeddings generating. For this task, we consider the co-authorship graph $\hat{G}(\hat{V}, \hat{A}, \hat{X})$, where $\hat{V} = \{\hat{v}_1, \hat{v}_2, \dots, \hat{v}_m\}$ is the set of graph vertexes (authors), $\hat{A}: m \times m \rightarrow \{0, 1\}$  is the adjacency matrix (each edge encodes collaboration between the two authors), and $\hat{X}: m \times k \rightarrow \textbf{R}$ denotes the matrix of node features (one-hot encoded research interests of authors). In order to supply the predictive model by additional data about publications of the authors, we extend each element $\hat{x}_i$ of the matrix $\hat{X}$ to $\hat{x}'_i$ as follows:  

\begin{equation}
    \hat{x}'_i = \operatorname{CONCAT} (x_i, \sum_{e\in e_i}  e),
\end{equation}
where $e_i$ denotes the set of publications embeddings of ${\rm i}^{th}$ author.  

\begin{figure}[t]
\includegraphics[width=\textwidth]{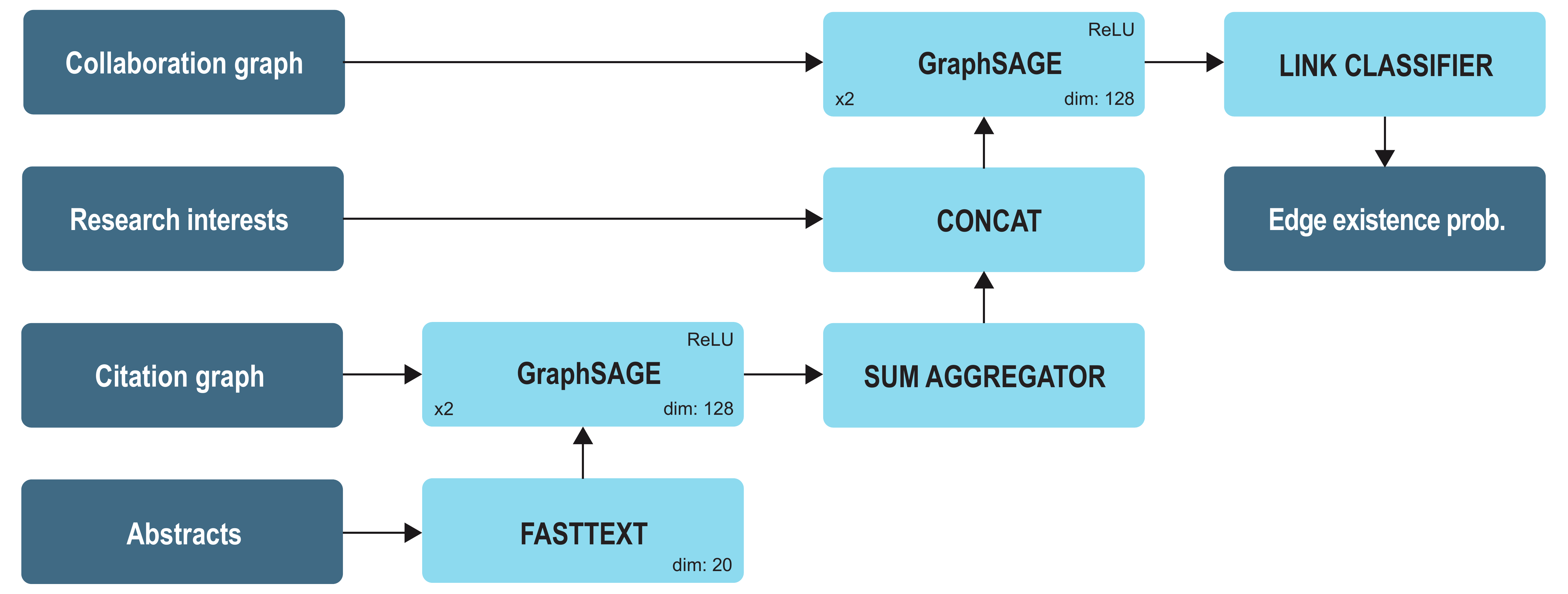}
\caption{Example configuration of the described architecture based on GraphSAGE convolutions.} 
\label{pipeline}
\end{figure}

After the concatenation, the extended graph was translated as an input to the two-layer GraphSAGE with link classifier. It constructs the embedding of the potential links applying a binary operator to the pair of node embeddings (we consider L1, L2, Hadamard operator, average, and inner product  \cite{link_embs_operators}). Finally, these link embeddings are passed through the dense classification layer to obtain probabilities of links existence in the network. The whole pipeline is illustrated in Figure \ref{pipeline}. 

The last model is trained by minimizing the binary cross-entropy loss 
\begin{equation}
L=-\frac{1}{\text { N }} \sum_{i=1}^{\text {N }} y_{i} \cdot \log \hat{y}_{i}+\left(1-y_{i}\right) \cdot \log \left(1-\hat{y}_{i}\right),
\end{equation}
where $N$ is the output size, $y_i$ is the true link labels, $\hat y_i$ is the predicted link existence probabilities. 

\section{Results}

We performed training and evaluation on the PC with 1 GPU Tesla V100 and 96 GB of RAM. The weights of the neural networks were updated by Adam optimizer. We divided graph edges into train, validation and test samples in ratio 3:1:2.

We evaluated the set of models with various link embedding operators, representation learning models, and aggregation functions. As the main quality measurements, we chose binary accuracy, AUC-ROC, and F1-score. Received embeddings from the models were tested on the supervised GraphSAGE setup applied to the link prediction task.

\begin{table}[t]
\begin{center}
\fontfamily{phv}\selectfont
\begin{tabular}{lccccc}
\rowcolor{LightCyan}
\rule{0pt}{4ex}
 Article embedding & Author embedding & LP op. & Accuracy & AUC-ROC & F1-score \hspace{2ex} \\[2ex]
\rule{0pt}{4ex}
-- & GraphSAGE (Mean) & L2 & 0.8793 & 0.9442  & 0.8817   \\
\rule{0pt}{4ex}  
FastText & GraphSAGE (Mean) & Had & 0.8844 & 0.9486 & 0.8828   \\
\rule{0pt}{4ex}  
GraphSAGE (Mean) & GraphSAGE (Mean) & Had & 0.8895 & 0.9568 & \textbf{0.8911}  \\
\rule{0pt}{4ex}  
GraphSAGE (Mean) & GraphSAGE (Mean) & L2 &  \textbf{0.8928} & 0.9531 & 0.8885 \\
\rule{0pt}{4ex}  
GraphSAGE (Mean) & GraphSAGE (MaxPool) & L1 & 0.8638 & \textbf{0.9617}  & 0.8489    \\
\end{tabular}
\end{center}
\caption{Results of different models on test sample}
\label{results_table}
\end{table}

We conducted experiments with more than 80 different configurations and represented key results in Table \ref{results_table}. The first two baselines were selected as the best among the approaches using either vectors of authors interests without citation network information or just embeddings of the abstracts. Comparison of these simpler architectures with full models which includes two graph neural networks could be interpreted as an ablation study.  

As it is shown in the table, the proposed aggregation algorithm positively influences the quality of scientific collaboration forecasting. The model without any citation data significantly suffers from the lack of expressive input features as well as the pipeline including only the abstracts of the papers. Obtained result allow us to report that proposed aggregation strategy efficiently utilizes citation graph properties. 

\section{Conclusion and Outlook}

In the present paper, we briefly introduced the two-stage pipeline for the collaboration prediction task. Performed computational experiments reveal the perspective of citation data utilization in sense of co-authorship network extension. The embeddings generated by GNNs effectively capture the network properties including its topology and vectorized abstracts represented as features of the corresponding graph nodes.

Our main contributions in the present work are the following:

\begin{enumerate}
    \item We perform extraction of the co-authorship graph from the corresponding HEP-TH citation network.
    \item Presence of useful metadata allows us to parse the scientific significance measures of the publications (e.g., impact factor).
    \item We aggregate structural data from the citation graph and apply it to the co-authorship network in order to evaluate its influence on the link prediction quality.
\end{enumerate}

Along with the future improvements of link prediction methods in the area of scientific collaboration, we intend to explore qualitative and quantitative assessment approaches of emerged links. As the probabilistic estimation of collaborations is not sufficient, it is important to extend it by less abstract metrics. In the following, we are going to leverage the average impact factor and the total number of publications for this task.

\section*{Acknowledgements}

We acknowledge fruitful discussions with Natalia Semenova.

%
%

\bibliographystyle{splncs04} 
\bibliography{main}
\end{document}